\documentstyle[aps,prb,twocolumn,floats,psfig]{revtex}

\begin{document}

\bibliographystyle{prsty}

\title{
%\begin{flushleft}
%{\small \em submitted to}\\
%{\small PHYSICAL REVIEW B \hfill $\qquad$ VOLUME
%{\normalsize XX},
%XX, NUMBER
%{\normalsize X}
%XX
%\hfill
%MONTH XX, YEAR XXXX }
% {\normalsize 1} MONTH {\normalsize YEAR}-I, {\normalsize
%startpage-endpage} }\\
%\end{flushleft}
Inertial mass of a superconducting vortex \vspace{-1mm} }

\author{
E. M. Chudnovsky$^1$ and A. B. Kuklov$^2$}

\address{ $^1$ Department of Physics and Astronomy, Lehman College,
City University of
New York, \\
250 Bedford Park Boulevard West, Bronx, New York 10468-1589
\\
$^2$ Department of Engineering Science and Physics, The College of
Staten Island, \\
City University of New York, Staten Island, New York 10314
%\date{\today}
%\maketitle
%\abstract{
\smallskip
%{\rm(Received DAY MONTH YEAR)}
\bigskip\\
%\parbox{14.2cm}
%\begin{abstract}
\parbox{14.2cm}
{\rm We show that a large contribution to the inertial mass of a
moving superconducting vortex comes from transversal displacements
of the crystal lattice. The corresponding part of the mass per
unit length of the vortex line is $M_{l} = ({\rm
m}_e^2c^{2}/64{\pi}{\alpha}^{2}{\mu}{\lambda}_{L}^{4})\ln({\lambda}_{L}/{\xi})$,
where ${\rm m}_{e}$ is the the bare electron mass, $c$ is the
speed of light, ${\alpha}=e^{2}/{\hbar}c\,{\approx}\,1/137$ is the
fine structure constant, ${\mu}$ is the shear modulus of the
solid, ${\lambda}_{L}$ is the London penetration length and
${\xi}$ is the coherence length. In conventional superconductors,
this mass can be comparable to or even greater than the vortex
core mass computed by Suhl.
%\end{abstract}
\smallskip
\begin{flushleft}
PACS numbers:  74.20.-z, 74.25.Kc, 74.25.Qt
\end{flushleft}
}}

\maketitle

In this Letter we analyze an old problem of the inertial mass of a
moving vortex line in a type-II superconductor. While in most
applications of superconductors the dynamics of vortex lines is
dissipative, the inertial mass can play an important role at slow
motion of vortices and in problems of quantum tunneling of
vortices. Besides, it is a fundamental problem of the theory of
superconductivity, not without a controversy, repeatedly addressed
by a number of researchers over the last forty years. We will show
that all previous works on this subject overlooked what appears to
be a significant contribution to the vortex mass in conventional
type-II superconductors: the torsional deformations of the crystal
lattice by the moving vortex. The brief history of the subject is
outlined below.

In 1965 Suhl [\onlinecite{Suhl}] computed two contributions to the
vortex mass: the mass due to the kinetic energy of the vortex
core, $M_{c}=(2/{\pi}^{3}){\rm m}k_{F}$, and the mass due to the
electromagnetic energy of the vortex,
$M_{em}=(r_{D}/{\xi})^{2}M_{c}\,$. Here ${\rm m}$ is the effective
electron mass, $k_{F}$ is the Fermi wave vector of normal
electrons, $r_{D}$ is the Debye charge-screening length, and
${\xi}$ is the superconducting coherence length. The vortex core
mass arises from the change in the effective mass of electrons by
~$({\Delta}/{\epsilon}_{F}){\rm m}$~ when one moves away from the
center of the core, with ${\Delta}$ and ${\epsilon}_{F}$ being the
superconducting gap and the Fermi energy respectively (see, e.g.,
Ref. [\onlinecite{Blatter}]). As to the electromagnetic mass, it
is simply due to the static energy of the magnetic flux carried by
the vortex (see, e.g., Ref. [\onlinecite{Leggett}]). The condition
${\xi} \gg r_{D}$ yields $M_{c} \gg M_{em}\,$. Suhl's results were
later reproduced by other methods (see, e.g., Refs.
[\onlinecite{Kupriyanov,Blatter,Leggett,Duan,Sonin}]). A separate
question, not addressed in this Letter, is the dynamic mass of the
vortex in a superclean limit which may be relevant to the
low-temperature physics of clean high-temperature superconductors.
This mass arises from the quantization of the electron states
inside the vortex core, which is important at small $\xi$. It was
shown to exceed the core mass by a large factor
$({\epsilon}_{F}/{\Delta})^{2}$,
[\onlinecite{Kopnin,Otterlo,Vinokur}].

In this Letter we will demonstrate that a moving vortex produces
torsional shear deformations of the crystal lattice that
contribute to the vortex mass. Such deformations are described by
the transversal displacement field ${\bf u}({\bf r}, t)$
satisfying
\begin{equation}\label{transversal}
{ \nabla}\cdot{\bf u}=0\;.
\end{equation}
In the long-wave limit they do not affect the density of the ionic
lattice.  Notice that the contribution to the vortex mass due to
longitudinal elastic deformations (${ {\nabla}}{\times}{\bf u}=0$)
has been studied before
[\onlinecite{Simanek,Coffey,Duan-Simanek}]. That contribution
comes from the tiny difference between the mass density of the
normal and superconducting phase and has nothing to do with the
effect studied by us. The latter effect arises from the
electromagnetic coupling between the superconducting current and
the local rotations of the crystal lattice.

We first recall a well-studied problem of a uniform rotation of a
superconductor at an angular velocity ${\bf \Omega}$. A simple
argument shows that the electron superfluid cannot remain at rest
when the ion lattice is rotating [\onlinecite{Anderson}]. That
would result in a huge current of ions relative to the electrons,
and the corresponding huge magnetic moment, ${\bf M}=-({\rm
e}/4c)NR^{2}{\bf \Omega}$, where ${\rm e}<0$ is the charge of
electron, $N$ is the macroscopic total number of superconducting
electrons, and $R$ is the radius of the cylinder. Obviously, this
is not the case. As the ions begin to move with respect to the
electron superfluid, the increasing magnetic flux due to the ionic
current induces, through Faraday's law, the eddy electric field
that drags superconducting electrons together with the rotating
ion lattice. Some effect of the rotation remains, however. It
comes from the Coriolis force on the electrons in the rotating
frame of reference (see, e.g., Ref. [\onlinecite{Cabrera}] and
references therein). Another way to look at it is that, in the
rotating frame, the electrons feel the effective magnetic field,
${\bf H}=(2{\rm m}_{e}c/{\rm e})\;{\bf \Omega}$. Consequently, due
to the Meissner effect, the rotating cylinder develops the
magnetic moment $4{\pi}{\bf M}=-(2{\rm m}_{e}c/{\rm e})\;{\bf
\Omega}$, so that the total field in the rotating frame, ${\bf
H}+4{\pi}{\bf M}$, is zero. This is a well-known effect predicted
by London [\onlinecite{London}] and first observed by Hildebrandt
[\onlinecite{Hildebrandt}]. Our goal is to study the effect of
local rotations,
\begin{equation}\label{Omega}
{\bf \Omega}({\bf r},t)=\frac{1}{2}{ \nabla}{\times}{\dot{\bf
u}}\;,
\end{equation}
caused by the transverse elastic deformations of the crystal
lattice.

A clear illustration of the problem in hand can be best achieved
within the Lagrangian formalism at $T=0$. Consider a simplified
model via London [\onlinecite{London}], in which each ion supplies
Z itinerant electrons, all of them superconducting at $T=0$. In
the long-wave limit the Lagrangian of the electronic superfluid
and the ionic lattice is
\begin{eqnarray}\label{Lagrangian}
{\cal{L}} & = & \int\,d^{3}r\,n\left[\frac{{\rm m}_e {\bf
v}_{L}^{2}}{2}+\frac{{\rm e}}{c}\,{\bf
v}_{L}{\cdot}{\bf A}\right] \nonumber \\
& + & \int\,d^{3}r\, \frac{n}{Z}\left[\frac{M_{i}{{\dot{\bf
u}}}^{2}}{2}-\frac{Z{\rm e}}{c}\,{\dot{\bf u}}{\cdot}{\bf A} -
{\lambda}_{iklm}u_{ik}u_{lm}\right].
\end{eqnarray}
Here $n$ and $n/Z$ are, correspondingly, the concentrations of
electrons and ions; $-Z{\rm e}\,>\,0$ and $M_{i}$ are the charge
and the mass of the ion, respectively; ${\bf v}_{L}(\bf r)$ is the
velocity of the superfluid in the laboratory coordinate frame,
${\bf A}$ is the vector potential, ${\lambda}_{iklm}$ is the
tensor of elastic coefficients, and
$u_{ik}=\frac{1}{2}({\partial}_{i}u_{k}+{\partial}_{k}u_{i})$ is
the strain tensor. The full Lagrangian is obtained by adding to
Eq.\ (\ref{Lagrangian}) the Lagrangian of the electromagnetic
field,
\begin{equation}\label{em}
{\cal{L}}_{em}=\frac{1}{8{\pi}}\,\int\,d^{3}r\,({\bf E}^{2}-{\bf
B}^{2})\,.
\end{equation}
Here ${\bf E}=-{\dot{\bf A}}/c$ is the eddy electric field,
satisfying ${ \nabla}{\cdot}{\bf E}=0$, $\;{ \nabla}{\times}{\bf
E}=-{\dot{\bf B}}/c$, and ${\bf B}={ \nabla}{\times}{\bf A}$ is
the magnetic field, satisfying ${ \nabla}{\times}{\bf
B}=(4{\pi}/c){\bf j}+{\dot{\bf E}}/c$, where ${\bf j}$ is the
superconducting current. If ${\bf j}$ depends on time, the term
$-(Z{\rm e}/c)\,{\dot{\bf u}}{\cdot}{\bf A}$ in Eq.\
(\ref{Lagrangian}) produces the force $-Ze{\bf E}$ in the equation
of motion for the ions that results in the elastic strain.

Since the electric current,
\begin{equation}\label{current}
{\bf j}=\frac{{\delta}{\cal{L}}}{{\delta}{\bf A}}={\rm e}n{\bf v}
= {\rm e}n({\bf v}_{L}-{\dot{\bf u}})\;,
\end{equation}
is determined by the motion of electrons with respect to the ions,
it is natural to switch in Eq.\ (\ref{Lagrangian}) to the variable
${\bf v} = {\bf v}_{L} - {\dot{\bf u}}\;$. Then Eq.\
(\ref{Lagrangian}) becomes
\begin{eqnarray}\label{Lagrangian-v}
{\cal{L}} & = & \int\,d^{3}r\,n\left[\frac{{\rm m}_e {\bf
v}^{2}}{2}+\frac{{\rm e}}{c}\,{\bf v}{\cdot}{\bf A} + {\rm m}_e{\bf v}{\cdot}{\dot{\bf u}}\right] \nonumber \\
& + & \int\,d^{3}r\, \frac{n}{Z}\left[\frac{(M_{i}+Z{\rm
m}_e){{\dot{\bf u}}}^{2}}{2}- {\lambda}_{iklm}u_{ik}u_{lm}\right].
\end{eqnarray}
The important feature of this equation is the term
$\int\,d^{3}r\,n{\rm m}_{e}{\bf v}{\cdot}{\dot{\bf u}}$ describing
the interaction between the electric current ${\bf j}=en{\bf v}$
and the dynamic transversal deformations of the ion lattice. We
see that ${\dot{\bf u}}$ plays the role similar to that of the
vector potential ${\bf A}$. Combining together the second and the
third term in Eq.\ (\ref{Lagrangian-v}), one can introduce the
effective vector potential seen by the electrons,
\begin{equation}\label{A-effective}
{\bf A}_{eff}  =  {\bf A}+ \frac{{\rm m}_{e}c}{\rm e}{\dot{\bf
u}}\;,
\end{equation}
and the corresponding effective magnetic field,
\begin{equation}\label{B-effective}
 {\bf B}_{eff} \; \equiv \; {
\nabla}{\times}{\bf A}_{eff} = {\bf B} + \frac{2{\rm m}_{e}c}{\rm
e}{\bf \Omega}\;,
\end{equation}
where ${\bf \Omega}({\bf r},t)$ given by Eq.\ (\ref{Omega}) is the
angular velocity of the local rotations. The second term in Eq.\
(\ref{B-effective}) is responsible for the London's magnetic
moment of a rotating superconductor discussed above.

In this Letter we want to study the effect of the transversal
displacements produced in the ion lattice by the electric field of
a moving fluxon. For that purpose, let us concentrate on the part
of the total action that contains ${\bf u}$,
\begin{equation}\label{Action}
S=\int dt \int d^{3}r\left[\frac{{\rm m}_{e}}{\rm e}\,{\bf
j}{\cdot}{\dot{\bf u}} + \frac{1}{2}{\rho}{{\dot{\bf u}}}^{2} -
{\mu}u_{ik}^{2}\right].
\end{equation}
Here ${\rho}=(n/z)M_{i}+n{\rm m}_{e}$ and ${\mu}$ are,
respectively, the total mass density and the shear modulus of the
solid which, for simplicity, we consider isotropic. The constant
${\mu}$ can be presented as ${\mu}={\rho}c_{t}^{2}$, where $c_{t}$
is the speed of the transverse sound. Notice that the elastic
energy of an isotropic solid also contains the term
$\frac{1}{2}{\lambda}u_{kk}^{2}$ with ${\lambda}+\frac{2}{3}{\mu}$
being the compression modulus [\onlinecite{LL-Elasticity}]. In our
case this term is zero due to the condition (\ref{transversal}).
According to Eq.\ (\ref{Action}), ${\bf u}$ satisfies
\begin{equation}\label{u}
\frac{d^{2}{\bf u}}{dt^{2}} -  c_t^2 \nabla^2 {\bf u}=- \frac{{\rm
m}_{e}}{\rm e \rho }\,\frac{d{\bf j}}{dt} \;.
\end{equation}

Looking at Eq.\ (\ref{u}), one makes a startling observation.
Since ${\bf j}$ is proportional to the electric charge, the charge
cancels out from Eq.\ (\ref{u}), as well as from Eq.\
(\ref{Action}), despite the electromagnetic nature of the forces
acting between the electronic and ionic subsystems. Eq.\
(\ref{u}), therefore, must be the consequence of the momentum
conservation rather than the result of a particular mechanism of
interaction between the particles of the crystal. This also
follows from the fact that the first term in Eq.\ (\ref{Action})
can be written in the Coriolis form, $\int d^{3}r\,{\bf
L}{\cdot}{\bf \Omega}$, where ${\bf L} =({\rm m}_{e}/{\rm e}){\bf
r}{\times}{\bf j}$ is the density of the angular momentum. The
latter is true, however, only if ${\rm m}_{e}$ and ${\rm e}$ in
Eq.\ (\ref{Lagrangian}) are bare electron mass and charge, which
was postulated by our simplified model. We shall now prove that
{\it in the low-frequency limit, Eq.\ (\ref{u}), with bare
electron mass and charge, is correct regardless of the model}.

We first notice that the electric current, expressed in terms of
the phase $\varphi$ of the superfluid wave function, is
\begin{equation}\label{j-GL}
{\bf j}=\frac{\hbar {\rm e}}{2{\rm m}}\,n\left( { \nabla} \varphi
-\frac{2{\rm e}}{\hbar c} {\bf A}_{eff}\right)\;,
\end{equation}
where ${\bf A}_{eff}$ is given by Eq.\ (\ref{A-effective}). While
(\ref{j-GL}) can be formally obtained from Eq.\ (\ref{Lagrangian})
in the spirit of the Ginzburg-Landau theory [\onlinecite{LP}], it
is in fact the most general form of the gauge invariant
superconducting current, with ~${\rm m}$~ being the effective
electron mass and ~$n$~ being the concentration of superconducting
electrons. A crucial observation in explaining the value of the
London's moment is that the electrons see the effective vector
potential of Eq.\ (\ref{A-effective}), with ${\rm m}_{e}$ and
${\rm e}$, being bare electron mass and charge. The latter follows
from the fact that equations (\ref{A-effective}) and
(\ref{B-effective}) are the consequence of the Larmore theorem
which provides the relation between ${\bf \Omega}$ and ${\bf B}$
(or, equivalently, the relation between the mechanical angular
momentum and the magnetic moment of electrons) in terms of the
bare electron mass and charge regardless of the interactions.

With Eq.\ (\ref{j-GL}) in mind, let us study the Lagrangian of the
system, ${\cal{L}}={\cal{L}}_{el}+{\cal{L}}_{f}$, expressed in
terms of coordinates of individual electrons ${\bf r}_{\alpha}$,
the phonon displacement field ${\bf u}$, and the electromagnetic
field ${\bf A}$. Here ${\cal{L}}_{el}$ includes all terms that
depend explicitly on ${\bf r}_{\alpha}$, while ${\cal{L}}_{f}$
describes long-wave deformations and electromagnetic fields.
Consider now the Raus function [\onlinecite{LL-Mechanics}]
\begin{equation}\label{Raus}
R=\sum_a {\bf p}_{\alpha}{\dot{\bf r}}_{\alpha} -
{\cal{L}}={\cal{H}}_{el} - {\cal{L}}_{f}\;,
\end{equation}
which is the Hamiltonian for electrons and (with minus sign) the
Lagrangian for ${\bf u}$ and ${\bf A}$. Here ${\bf p}_{\alpha}
=\partial {\cal{L}}/\partial {\dot{\bf r}}_{\alpha}$ are canonical
electron momenta. Because electrons adiabatically adjust to the
low-frequency long-wavelength displacements of the lattice and the
electromagnetic field, one can quantize ${\cal{H}}_{el}$ at any
point in a solid as if ${\bf A}$ and ${\bf u}$ were constant.
Since the electrons see the effective vector potential of Eq.\
(\ref{A-effective}), the resulting free energy, $F$, must depend
on ${\bf u}$ and ${\bf A}$ through the combination
(\ref{A-effective}), $F=F[{\bf A}_{eff}]$. The gauge invariant
superconducting current must now be obtained as
[\onlinecite{LL-Quantum}]
\begin{equation}\label{j-F}
{\bf j}=-c\,\frac{\delta F[{\bf A}_{eff}]}{\delta {\bf
A}}=-\frac{{\rm e}}{{\rm m}_{e}}\,\frac{\delta F[{\bf
A}_{eff}]}{\delta \dot{\bf u}} \;.
\end{equation}
Consequently, the effective Lagrangian for ${\bf u}$ and ${\bf A}$
is
\begin{equation}\label{L-eff}
{\cal{L}}_{eff}=-R_{eff}={\cal{L}}_{f} - F[{\bf
A}_{eff}]={\cal{L}}_{f}+\frac{{\rm m}_e}{\rm e}\int {\bf
j}{\cdot}d\dot{\bf u}\,.
\end{equation}
The part of this Lagrangian that depends on ${\bf u}$ determines
the effective action for the deformations,
\begin{equation}\label{Action-eff}
S_{eff}[{\bf u}]=\int dt \int d^{3}r\left[\frac{{\rm m}_{e}}{\rm
e}\int {\bf j}{\cdot}d{\dot{\bf u}} + \frac{1}{2}{\rho}{{\dot{\bf
u}}}^{2} - {\mu}u_{ik}^{2} \right].
\end{equation}
The variation of $S_{eff}[{\bf u}]$ on ${\bf u}$ gives Eq.\
(\ref{u}), which concludes the formal proof of that equation for
arbitrary interactions between the particles of the solid. {\it
Eq.\ (\ref{u}) is exact as long as the lattice deformations are
sufficiently slow to allow for local thermodynamic equilibrium in
the electronic subsystem}.

We shall now proceed with the computation of the effective mass of
a moving vortex. According to Eq.\ (\ref{u}), such a vortex
produces torsional deformations of the crystal. If the magnetic
flux was not quantized, the full treatment of the problem would
require a self-consistent solution of the equations of motion for
the deformation field ${\bf u}$ and the vortex current ${\bf j}$.
The quantization of the flux, however, makes the vortex a robust
source of deformations in the right-hand-side of Eq.\ (\ref{u}).
Let the vortex line be oriented along the Z-axis. Then, the
current depends on coordinates and time through ${\bf j}({\bf
r}-{\bf V}t)$, where ${\bf r}$ is the radius-vector perpendicular
to $Z$. For the vortex moving at a speed $V \ll\ c_{t}$, the terms
in the action and in the equation of motion, that are quadratic on
$\dot{\bf u}$, have a $(V/c_{t})^{2}$ smallness compared to other
terms and can be neglected. In that limit, Eq.\ (\ref{Action}) and
Eq.\ (\ref{Action-eff}) are equivalent. Integrating the first term
in Eq.\ (\ref{Action}) by parts and omitting small terms, we get
\begin{equation}\label{Action-byparts}
S=\int dt \int d^{3}r\left[\frac{{\rm m}_{e}}{\rm e}\,{\bf
u}{\cdot}({\bf V}{\cdot}{ \nabla})\,{\bf j} -
{\mu}u_{ik}^{2}\right].
\end{equation}
Since both ${\bf V}$ and ${\bf j}$ are perpendicular to $Z$, it is
clear that the spatial Fourier harmonics of ${\bf u}$ that couple
to ${\bf j}$ must have ${\bf k}$ and ${\bf u}$ also perpendicular
to $Z$. The problem then becomes essentially two-dimensional as,
of course, is expected from its symmetry. In terms of the Fourier
transforms of ${\bf j}({\bf r})$ and ${\bf u}({\bf r})$ one
obtains
\begin{eqnarray}\label{Action-2D}
S & = & \frac{1}{2}\int dt \int dz \int
\frac{d^{2}k}{(2{\pi})^{2}}\, [- {\mu}k^{2}{\bf u}_{\bf
k}{\cdot}{\bf u}_{\bf k}^{*} \nonumber \\
& + & \frac{{\rm m}_{e}}{\rm e}\,i\,({\bf k}{\cdot}{\bf V})\,({\bf
j}_{\bf k}{\cdot}{\bf u}_{\bf k}^{*} - {\bf j}_{\bf
k}^{*}{\cdot}{\bf u}_{\bf k}) ]\;.
\end{eqnarray}
The minimization of this action on ${\bf u}_{\bf k}$ is equivalent
to taking the Fourier transform of Eq.\ (\ref{u}) in the limit of
small $V$. Substitution of the solution of that equation back into
Eq.\ (\ref{Action-2D}) yields
\begin{eqnarray}\label{Action-final}
S & = & \int dt \int dz \; \frac{M_{l}V^{2}}{2} \nonumber
\\
& = & \frac{{\rm m}_{e}^{2}}{2{\rm e}^{2}{\mu}}\,\int dt \int dz
\int \frac{d^{2}k}{(2{\pi})^{2}}\,\frac{({\bf k}{\cdot}{\bf
V})^{2}}{k^{2}}{\bf j}_{\bf k}{\cdot}{\bf j}_{\bf k}^{*} \; .
\end{eqnarray}
Further averaging over the angles gives for the vortex mass per
unit length along the Z-axis:
\begin{equation}\label{mass-current}
M_{l} = \frac{{\rm m}_{e}^{2}}{2{\rm e}^{2}{\mu}} \int d^{2}r \,
j^{2}(r) \, ,
\end{equation}
where $j(r)$ is the current of a stationary vortex.

The integral in Eq.\ (\ref{mass-current}) can be expressed in
terms of the unit-length energy of the vortex line, $E_v$,
[\onlinecite{LP}]. In the limit of ${\kappa}={\lambda}_{L}/\xi \gg
1$,
\begin{equation}\label{vortex-energy}
E_{v} = \frac{2{\pi}{\lambda}_{L}^{2}}{c^{2}}\int d^{2}r \,
j^{2}(r) =
\left(\frac{{\Phi}_{0}}{4{\pi}{\lambda}_{L}}\right)^{2}\ln{\kappa}
\; ,
\end{equation}
where ${\Phi}_{0}=ch/2{\rm e}$ is the flux quantum and
${\lambda}_{L}$ is the London penetration length. This gives for
the vortex mass per unit length
\begin{equation}\label{mass-final}
M_{l} = \frac{{\rm
m}_{e}^{2}c^{2}}{64{\pi}{\alpha}^{2}{\mu}{\lambda}_{L}^{4}}\ln{\kappa}
= \frac{\pi}{4} \frac{({\hbar}n)^{2}}{\mu}\left(\frac{{\rm
m}_{e}}{\rm m}\right)^{2}\ln{\kappa} \;,
\end{equation}
where we introduced the fine structure constant,
${\alpha}=e^{2}/{\hbar}c\,{\approx}\,1/137$, and used the relation
${\lambda}_{L}^{2}={\rm m}c^{2}/4{\pi}n{\rm e}^{2}$ to obtain the
second of Eq.\ (\ref{mass-final}).

Three observations are in order. Firstly, the non-zero value of
$M_{l}$ is due to the finite rigidity of the crystal with respect
to the shear stress. In an absolutely rigid crystal,
${\mu}={\rho}c_{t}^{2}=\infty$ and $M_{l}$ would be zero.
Secondly, ${M}_{l}$ scales as the square of the superfluid
density, which is a rather unique feature provided by the specific
mass-generating mechanism studied in this Letter. Consequently,
$M_{l}$ should go to zero as $(T_{c}-T)^{2}$ when temperature
approaches the critical temperature $T_{c}$. Thirdly, $M_{l}$ does
not fall into the category of vortex masses [\onlinecite{Leggett}]
that satisfy $M_{i}\,{\sim}\,E_{i}/c_{i}^{2}$, where $E_{i}$ is a
contribution of some mechanism to the energy of a static vortex
and $c_{i}$ is the velocity of propagation of the distortion in
question. This is because no static energy is associated with the
mechanism that generates $M_{l}$.

Eq.\ (\ref{mass-final}) shows that the vortex mass studied in this
Letter is important in metals with high concentration of
superconducting electrons. In copper oxides, because of low $n$,
and also due to strong dynamic effects associated with the
quantization of the electron levels in the vortex core
[\onlinecite{Kopnin,Otterlo,Vinokur}], this mass should not play
any significant role. In good metals, however, $M_{l}$ can be the
main contribution to the inertial mass of the vortex line. Taking
for estimates, $n\,{\sim}\,10^{23}\,$cm$^{-3}$,
$\;{\mu}\,{\sim}\,10^{11}\,$g/cm$^{3}$, and ${\rm m}\,{\sim}\,{\rm
m}_{e}$, one gets from the second of Eq.\ (\ref{mass-final}),
$M_{l}\,{\sim}\,10^{-19}\,$g/cm. Thus, $M_{l}$ can easily exceed
the inertial mass of the vortex core computed by Suhl,
$M_{c}=(2/{\pi}^{3}){\rm m}k_{F}$
[\onlinecite{Suhl,Kupriyanov,Blatter}].
It is interesting to note, that, within the model of an isotropic
good metal, these two masses do not have any parameter smallness
with respect to each other. Indeed, taking at $T=0$,
$\;n=k_{F}^{3}/3{\pi}^{2}$, $\;{\mu}={\rho}c_{t}^{2}$,
$\;{\rho}=M_{i}n/Z$, we get
\begin{equation}\label{ratio}
\frac{M_l}{M_c}=\frac{{\pi}^{2}}{8}\,\left(\frac{c_{l}}{c_{t}}\right)^{2}\,\left(\frac{{\rm
m}_{e}}{\rm m}\right)^{2}\,\ln{\kappa}\;,
\end{equation}
where we have used the plasma approximation for the speed of the
longitudinal sound, $c_{l}=(Z{\rm
m}/3M_{i})^{1/2}v_{F}\,>\,\sqrt{2}c_{t}$
[\onlinecite{LL-Elasticity}]. However, when using the parameters
of real materials, $M_{l}$ seems always to exceed $M_{c}$ by
one-two orders of magnitude.

Finally, we would like to comment on the contribution of the
interaction term in equations (\ref{Action}),(\ref{Action-eff}) to
the viscosity for the vortex motion. Conservation of linear
momentum prohibits the radiation of transversal phonons by a
vortex moving at $V\,<\,c_{t}$. At $V\,>\,c_{t}$, however, and
especially close to $c_{t}$, the viscosity for the vortex motion
may, in principle, be dominated by the radiation of the
transversal sound rather than by the conventional mechanism due to
the finite normal-state resistivity [\onlinecite{Bardeen}]. The
full solution of this problem will be presented elsewhere.

In summary, we have computed the contribution to the inertial mass
of a moving superconducting vortex, which is coming from the
torsional deformations of the crystal lattice. Rigorous solution
of this problem for an isotropic solid has been obtained. This
contribution to the vortex mass can dominate over all other
contributions in metals with high concentration of superconducting
electrons.

This work has been supported by the U.S. DOE Grant No.
DE-FG-2-93ER45487.
%

%\vspace{-0.5cm}

%\bibliography{gar}

\end{document}